\documentclass[12pt]{article}
\pdfoutput=1
\usepackage{graphicx}
\usepackage{hyperref}
\usepackage{epsfig}
\usepackage{amsmath}
\usepackage{amssymb}
\usepackage{color}
\usepackage{cite} 
\usepackage{hyperref}
\usepackage{amssymb}
\usepackage[fontsize=11.45pt]{scrextend}
\usepackage{braket,mathrsfs,slashed}
\usepackage{setspace}
\newlength{\abstractwidth}
\renewcommand{\title}[1]{\vbox{\center\bf{\Large{#1}}}\vspace{5mm}}
\renewcommand{\author}[1]{\vbox{\center#1}\vspace{5mm}}
\newcommand{\address}[1]{\vbox{\center\em#1}}

\usepackage{dsfont}
\usepackage[paper=a4paper,margin=1.5cm]{geometry}
\usepackage[utf8]{inputenc}
\usepackage{rotating}
\usepackage{soul}

\usepackage{mathrsfs} 
\usepackage{bbm}
\usepackage{etoolbox} 

\usepackage[caption=false,lofdepth,lotdepth]{subfig}

\hypersetup{pdfstartview=FitV,colorlinks=true,linkcolor=midblue,citecolor=midblue,filecolor=midblue,urlcolor=midblue}

\definecolor{midblue}{rgb}{0,0,0.5}

\begin{document}
		\newgeometry{top=3cm,bottom=1cm,right=1.62cm,left=1.62cm}

	\begin{titlepage}
\begin{flushright} {\footnotesize YITP-21-76, IPMU21-0049}  \end{flushright}
		\begin{center}
			\hfill \\
			\hfill \\
			\hfill \\
			\vskip 0.5cm

			\title{\Large On the assumptions leading to the information loss paradox}

			\author{\large Luca Buoninfante$^{a,\,\star}$, Francesco Di Filippo$^{b,\,\dagger}$ and Shinji Mukohyama$^{b,c,\,\ddagger}$
						}
			
				\address{$^a$Department of Physics, Tokyo Institute of Technology, Tokyo 152-8551, Japan.\\
				\vspace{.2cm}
				$^b$Center for Gravitational Physics, Yukawa Institute for Theoretical Physics,\\
				Kyoto University, 606-8502, Kyoto, Japan.\\
				\vspace{.2cm}

				$^c$Kavli Institute for the Physics and Mathematics of the Universe (WPI),\\
				The University of Tokyo, Kashiwa, Chiba 277-8583, Japan.
						}
		\end{center}

		\begin{abstract}
        The information loss paradox is usually stated as an incompatibility between general relativity and quantum mechanics. However, the assumptions leading to the problem are often overlooked and, in fact, a careful inspection of the main hypothesises suggests a radical reformulation of the problem. 
        Indeed, we present a thought experiment involving a black hole that emits radiation and, independently of the nature of the radiation, we show the existence of an incompatibility between (i)~the validity of the laws of general relativity to describe infalling matter far from the Planckian regime, and (ii)~the so-called central dogma which states that 
        as seen from an outside observer a black hole behaves like a quantum system whose number of degrees of freedom is proportional to the horizon area.
        We critically revise the standard arguments in support of the central dogma, and argue that they cannot hold true unless some new physics is invoked even before reaching Planck scales. 
        This suggests that the information loss problem, in its current formulation, is not necessarily related to any loss of information or lack of unitarity.
        Therefore,  in principle, semiclassical general relativity and quantum mechanics can be perfectly compatible before reaching the final stage of the black hole evaporation where, instead, a consistent theory of quantum gravity is needed to make any prediction.
		\end{abstract}
			\vspace{6.6cm}
			\noindent\rule{7.5cm}{0.5pt}\\
			$\,^\star$
			\href{mailto:buoninfante.l.aa@m.titech.ac.jp}{buoninfante.l.aa@m.titech.ac.jp}\\	
			$\,^\dagger$ \href{mailto:francesco.difilippo@yukawa.kyoto-u.ac.jp}{francesco.difilippo@yukawa.kyoto-u.ac.jp}\\
			$\,^\ddagger$ \href{mailto:shinji.mukohyama@yukawa.kyoto-u.ac.jp}{shinji.mukohyama@yukawa.kyoto-u.ac.jp}
	\end{titlepage}

	\newgeometry{top=3cm,bottom=4cm,right=2.5cm,left=2.5cm}

	{
		\hypersetup{linkcolor=black}
		\tableofcontents
	}
	
	\baselineskip=17.63pt 
	
	\newpage
	
	

\section{Introduction}
	
Observations of black hole exteriors constitute one of the most successful confirmation of general-relativity predictions~\cite{Abbott:2016blz}. On the other hand, black hole interiors are still puzzling and  lack a consistent theory capable of describing them due to the presence of a singularity where Einstein’s theory is expected to break down~\cite{Penrose:1964wq}. 
According to classical general relativity, this would not be a problem for outside observers as a black hole is endowed with a horizon that hides the singularity. However, at the semiclassical level -- i.e. with matter fields quantized in a classical background -- black holes turn out to be less dark than their classical counterpart, indeed they evaporate by emitting radiation whose spectrum is approximately thermal~\cite{Hawking:1974sw}. 
This feature has puzzled physicists for a long time by raising ``apparent'' contradictions. The most famous is the ``information loss paradox'' according to which an initial pure state describing collapsing matter would evolve into a final mixed state after black hole evaporation, thus giving rise to an inconsistency between general relativity and quantum mechanics as \textit{no} unitary operator exists that can evolve pure states into mixed states~\cite{Hawking:1976ra}; see also Ref.~\cite{Mathur:2009hf} for a pedagogical review.

It should be emphasized that this problem appears because one tries to extend the validity of the semiclassical approach up to the Planck scale, and extrapolate physical predictions in regimes where quantum-gravity effects should not be negligible. In this respect, there is \textit{not} really a paradox but it could be just our ignorance about Planck-scale physics that prevents us from concluding anything definite about what happens to the  information during black hole evaporation~\cite{Unruh:2017uaw}.

A different version of the paradox involving time scales shorter than black hole life time was formulated. Indeed, if one makes the ``apparently'' natural assumption that as seen from the outside a black hole behaves like a quantum system whose number of degrees of freedom is given by the horizon area, then one can show that  after some time the von Neumann entropy of Hawking radiation would exceed the thermodynamic entropy of black hole, which is impossible. This gives rise to a contradiction even before reaching the end point of the evaporation when the semiclassical approach should still be valid. Thus, it was argued that the von Neumann entropy of the Hawking radiation should start decreasing at roughly half of the black hole life time, and follow the so-called Page curve consistently with the unitary quantum evolution~\cite{Page:1993df,Page:1993wv,Page:1993up}.  

In the past decades both the aforementioned starting assumption and the corresponding alternative version of the information loss problem have been taken very seriously by a large part of the theoretical physics community. To emphasize the relevance of the hypothesis on the black hole degrees of freedom as seen from an outside observer, some authors even referred to it with the expression ``central dogma''~\cite{Almheiri:2020cfm}. Indeed, lots of efforts both to support the central dogma and to find a resolution to the induced entropy problem have been made. 
In particular, recently a novel computation~\cite{Penington:2019npb,Penington:2019kki,Almheiri:2019psf,Almheiri:2019hni,Almheiri:2019qdq,Marolf:2020rpm,Bousso:2021sji} of the von Neumann entropy of the Hawking radiation by using the Ryu-Takayanagi entropy formula~\cite{Ryu:2006bv} was shown to be consistent with a unitarity quantum evolution, and was claimed to be a strong motivation in support of the central dogma.
    
In this paper we adopt an open-minded and critical point of view in discussing whether the current formulation of the information loss paradox is well-posed. We aim at carefully inspecting all the assumptions that are usually made and that are claimed to lead to the information loss problem(s). In particular, in Sec.~\ref{Sec:info paradox} we review various formulations of the problem by emphasizing some of the details that are often overlooked, and that are necessary for the subsequent analysis. In Sec.~\ref{Sec:arguments} we examine the standard arguments in favour of the central dogma. In Sec.~\ref{Sec:counterex}, we present a thought experiment which unveils an incompatibility between the validity of semiclassical gravity to describe infalling matter far from the Planckian regime and the central dogma.  This allows us to argue that as long as the low-energy effective field theory holds true, i.e. far from the Planck scale, such that the horizon can be described as a smooth surface, then there exists \textit{no} information loss paradox. Moreover, in Sec.~\ref{Sec:micr} we comment on the microscopic interpretation of the Bekenstein-Hawking entropy; while in Sec.~\ref{Sec:Comp-fir} we discuss the relevance of our conclusions in comparison to the black hole complementarity and the firewall paradigm. In Sec.~\ref{Sec:discussions}, we summarize our results and discuss the power of our conclusions. 
    
Throughout the work we adopt the units $c=\hbar=k_{\rm B}=1.$ We will use different symbols for several entropies: $S_{\rm BH}$ for the Bekenstein-Hawking entropy; $S_{\rm bh}$ for the total black hole entropy that does not necessarily coincide with $S_{\rm BH}$; $S_{\rm th}$ for the thermodynamic entropy; $S_{\rm vN}$ for the von Neumann entropy; $S_{\rm rad}$ for the von Neumann entropy of the Hawking radiation; $S_{\rm m}$ for the von Neumann entropy of additional matter.

	\section{Is there an information loss paradox?}\label{Sec:info paradox}
	
	It is often stated that the thermal nature of black holes eventually leads to an incompatibility between semiclassical general relativity and one of the pillar of quantum mechanics, i.e. the unitarity of the $S$-matrix, thus leading to the well-known information loss paradox~\cite{Hawking:1976ra}. However, we strongly believe that the assumptions on which 
	the paradoxical implications rely are often overlooked, and deserve a deeper and thorough inspection.
	Therefore, it is crucial to reformulate the problem by highlighting and critically revising all the basic assumptions that are usually made.  In what follows we carefully analyze three different scenarios that are normally considered.

	\subsection{Case A: unitarity problem}
	
\begin{figure}[t]
	\centering
	\subfloat[Subfigure 1 list of figures text][]{
		\includegraphics[scale=0.42]{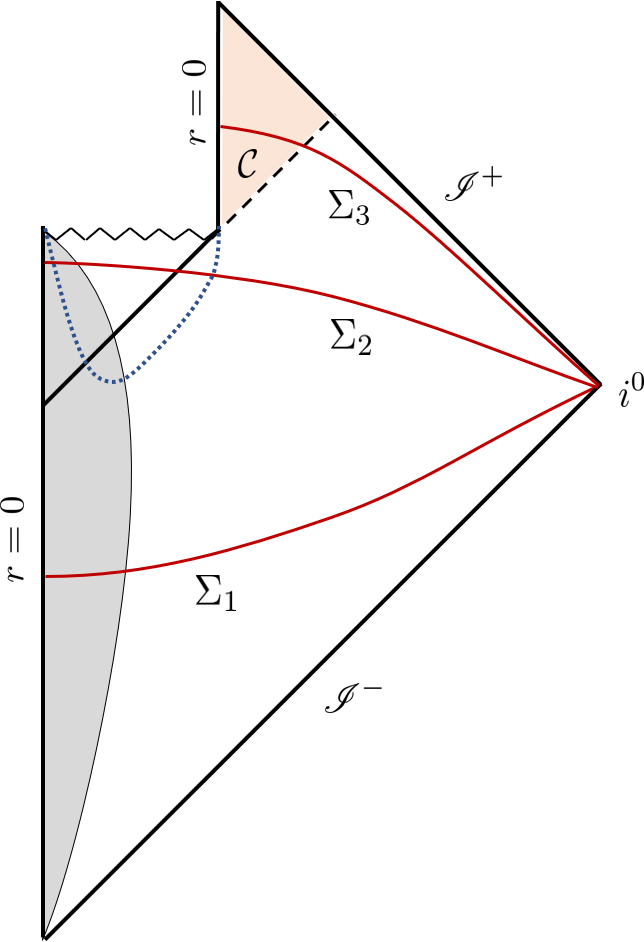}\label{fig1a}}\qquad
	\subfloat[Subfigure 2 list of figures text][]{
		\includegraphics[scale=0.42]{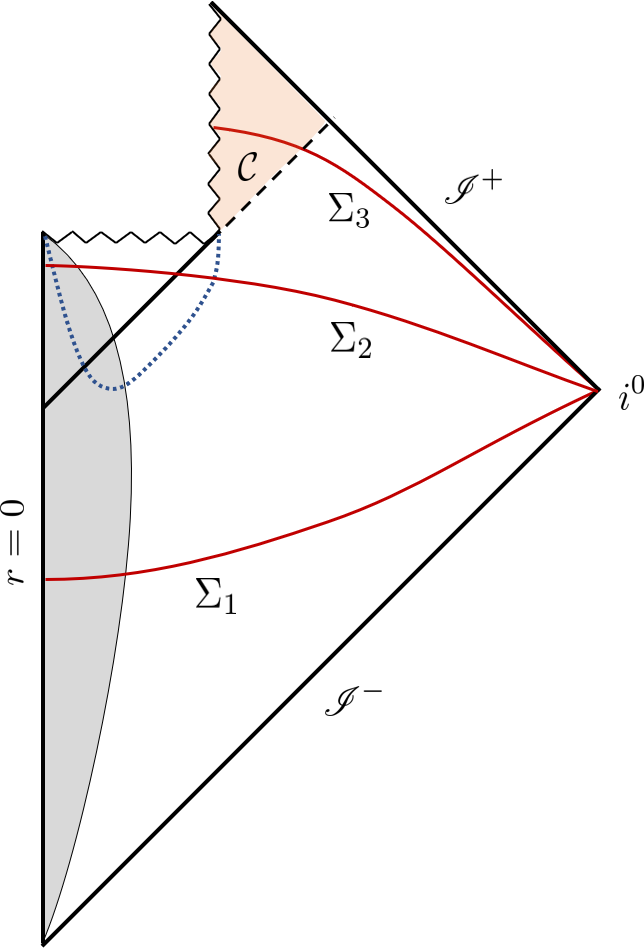}\label{fig1b}}\qquad
		\subfloat[Subfigure 3 list of figures text][]{
		\includegraphics[scale=0.42]{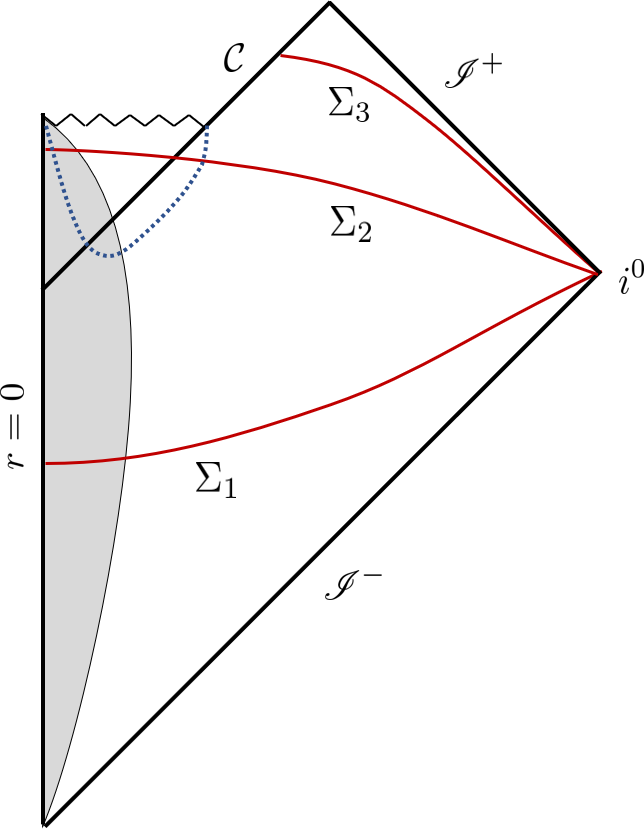}\label{fig1c}}
	\protect\caption{Penrose diagrams of an evaporating black hole with (a) regular and (b)  singular spacetime (naked timelike singularity) as the final state. In both cases, the spacetime has a Cauchy horizon labeled by $\mathcal{C}$. If the final state is singular or we only consider the maximal Cauchy development (c), a unitary operator does \textbf{not} necessarily evolve a pure state on $\Sigma_2$ into a pure state on $\Sigma_3$. The blue dotted line represents the ($\text{inner}\,\cup\, \text{outer}$) trapping horizon.}
\end{figure}

	The simplest formulation of the information loss problem can be stated as the incompatibility between the two following assumptions.
	\begin{enumerate}
		\item[A1.] Quantum states evolve in a unitary way. In particular, pure states evolve into pure states;
		
		\item[A2.] Semiclassical general relativity is a valid low-energy effective field theory to describe black hole physics during the entire evaporation process:
		black holes evaporate completely emitting thermal radiation and end up leaving a regular spacetime (see Fig.~\ref{fig1a}). 
	
    \end{enumerate}
	The hypothesis A1 implies that there exists a unitarity $S$-matrix operator that describes the evolution from ingoing collapsing matter to outgoing radiation during the entire black hole formation and evaporation. Whereas A2 means that the semiclassical approach, according to which fields are quantized in a classical curved background, is always valid even at the end point of the evaporation.
	
	These two assumptions are clearly incompatible. If the emitted radiation is in a pure thermal state (as suggested by Hawking's calculation~\cite{Hawking:1974sw}), then  after the black hole evaporation the final state will be thermal and mixed. This scenario is illustrated in Fig.~\ref{fig1a}, in which it is clear that before reaching the end point of the evaporation we can draw Cauchy surfaces, e.g. $\Sigma_1$ and $\Sigma_2,$ on which the state of the joint system (black hole plus radiation) is pure; whereas after the evaporation the state on each Cauchy slice turns out to be mixed, e.g. on $\Sigma_3.$
	Therefore, if the matter that formed the black hole was initially in a pure state, the previous argument would imply a breakdown of the unitary condition on the evolution operator in quantum mechanics as pure states cannot evolve unitarily into mixed states, thus contradicting A1.

	Obviously, there is no reason to trust the semiclassical picture up to the end of the black hole evaporation, indeed it is quite reasonable to believe that at least in the latest stages quantum-gravity effects cannot be neglected and should be consistently taken into account. In fact, it is \textit{not} clear what the final state would be. For instance, it could be given by a naked timelike singularity~\cite{Hawking:1974sw} as depicted in Fig.~\ref{fig1b}. 
	If this is the case, we would not be able to conclude that the final state is mixed. Indeed, $\Sigma_3$ in Fig.~\ref{fig1b} is not a Cauchy hypersurface, meaning that we would need a description of the singularity in order to predict the final quantum state.

	Furthermore, let us note that both the regular spacetime and the naked singularity spacetime represent only some arbitrary extrapolation of semiclassical gravity beyond its regime of validity. In fact, these spacetimes are not globally hyperbolic and contain a Cauchy horizon $\mathcal{C}$. The shaded orange regions in Figs.~\ref{fig1a} and~\ref{fig1b} are in the causal future of the singularity, and it is not possible to evolve the initial value problem from the hypersurface $\Sigma_2$ to $\Sigma_3$ without providing a description of the singularity which is outside the realm of semiclassical gravity. Semiclassical gravity can only study the maximally Cauchy development \cite{Wald:1984rg} depicted in Fig.~\ref{fig1c}. In this portion of spacetime the initial value problem is well defined. However, the spacetime is not geodesically complete, therefore in principle we should not expect a pure state on $\Sigma_2$ to evolve into a pure state on $\Sigma_3.$ Although in a different language, similar considerations were presented in \cite{Unruh:2017uaw}.

	Hence, this weak formulation of the information loss problem would not be particularly worrisome.  On the other hand, it is well known that one can formulate a stronger version of the paradox according to which problems seem to arise even when the black hole mass is much larger than Planck mass.


	\subsection{Case B: entropy problem}\label{CaseB}

	We now present a stronger formulation of the information problem originally due to Page~\cite{Page:1993df,Page:1993wv,Page:1993up}, that it is sometimes referred to as the ``entropy problem''; see also Ref.~\cite{Almheiri:2020cfm}. It consists in the incompatibility among the following assumptions.
	\begin{enumerate}
		
		\item[B1.] Quantum states evolve in a unitary way. In particular, pure states evolve into pure states;
		
		\item[B2.] Semiclassical general relativity is a valid low-energy effective field theory to describe black hole physics far from the Planckian regime;
		
		\item[B3.] As seen from the outside, a black hole behaves like a quantum system whose number of degrees of freedom is given by $A/4G,$ with $A$ being the apparent-horizon area.
		
	\end{enumerate}
	It is clear that in this formulation the first assumption is left unchanged  ($\text{A1}=\text{B1}$), whereas the second one has been significantly weakened ($\text{B2}\subset \text{A2}$). Indeed, B2 means that the semiclassical approach can be trusted only for black holes whose mass is sufficiently larger than Planck mass, and the horizon is still assumed to be a smooth surface, i.e. an infalling observer would experience nothing special when crossing the horizon, consistently with the equivalence principle.
	
    Furthermore, it is important to note that, if the mass of the black hole is sufficiently larger than Planck mass, it is possible to foliate the spacetime in a way that each Cauchy hypersurface intersects all the infalling matter and outgoing Hawking radiation, but avoids regions with high curvature. Also, the extrinsic curvature of the hypersurfaces remains small. This foliation is usually referred to as ``nice slicing''. As a consequence of the nice slicing, if B2 holds, the semiclassical description will be valid also for the matter and Hawking particles falling into the black hole.
	
	The price to pay in this formulation of the information problem is the introduction of the third hypothesis B3 that is sometime referred to as ``central dogma''~\cite{Almheiri:2020cfm}, according to which the black hole entropy\footnote{Here, we consider the black hole entropy as the number of physical degrees of freedom of a black hole as seen from the outside.} $S_{\rm bh}$ simply coincides with the Bekenstein-Hawking entropy $S_{\rm BH}=A/4G,$ i.e. $S_{\rm bh}=S_{\rm BH}$~\cite{Bekenstein:1972tm,Bekenstein:1973ur}; in other words the dimension of the black hole Hilbert space is assumed to be $e^{S_{\rm BH}}.$  We will discuss the motivations and plausibility of the central dogma in the next section. Before that, let us quickly repeat the standard argument~\cite{Page:1993wv,Page:1993up,Almheiri:2020cfm} according to which requiring the simultaneous validity of assumptions B1, B2, B3 leads to a paradoxical conclusion.

\begin{figure}[t!]
	\includegraphics[scale=0.55]{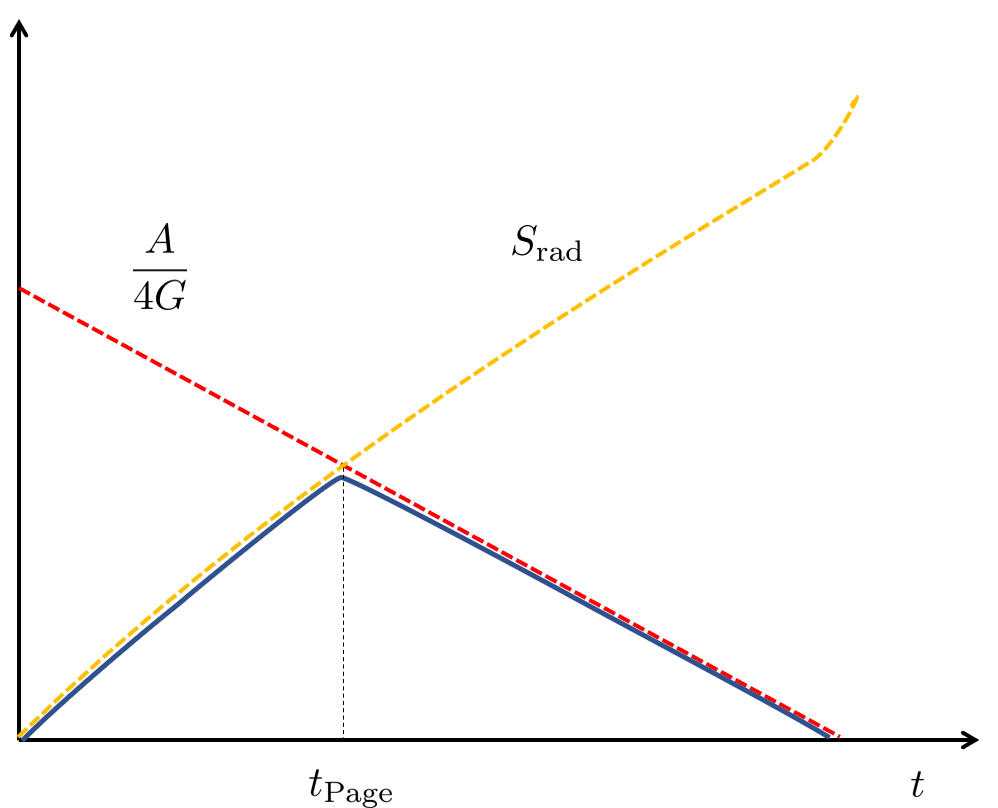}
	\centering
	\protect\caption{This figure shows the behaviors of Bekenstein-Hawking entropy $A/4G$ (red dashed line), and of the von Neumann entropy of the Hawking radiation $S_{\rm rad}$ (yellow dashed line) as functions of the time $t$ defining the foliation of Cauchy hypersurfaces. If the thermodynamic entropy of black hole $S_{\rm bh}$ is equal to $A/4G$ (i.e. if B3 holds), then for times larger than the Page time $t_{\rm Page}$ the entropy of radiation exceeds the thermodynamic one and leads to the information loss problem. In this scenario, the issue would be solved if and only if the entropy of radiation would follow the so-called Page curve (blue solid line).}
	\label{fig3}
\end{figure}

Let $\mathcal{H}_{\rm bh}$ and $\mathcal{H}_{\rm rad}$ be the Hilbert spaces of the black hole and of the radiation, respectively. B1 implies that, if the initial state before black hole formation was pure, then the joint state of black hole plus radiation, i.e. $\left|\psi \right\rangle \in \mathcal{H}_{\rm bh}\otimes \mathcal{H}_{\rm rad},$ must also be pure. By tracing over the black hole degrees of freedom, we can obtain the density matrix for the radiation subsystem and its von Neumann entropy $S_{\rm rad}$ (which coincides with the von Neumann entropy of the black hole subsystem). From B3 it follows that during the evaporation process the thermodynamic black hole entropy $S_{\rm bh}$ decreases as the horizon area decreases. Whereas the entropy $S_{\rm rad}$ of the outgoing radiation tends to increase according to Hawking semiclassical computation which can be trusted as long as B2 is valid. This means that there exists a time scale\footnote{Note that different notions of time are possible. We will use the time function orthogonal to the Cauchy foliation $\Sigma_t$ (each Cauchy hypersurface is a constant-time hypersurface).} $t_{\rm Page}$ -- known as Page time -- after which $S_{\rm rad}>A/4G=S_{\rm BH}=S_{\rm bh};$ see Fig.~\ref{fig3} for an illustration. 
The last inequality states that the von Neumann entropy of the Hawking radiation (or, equivalently, of the black hole subsystem) becomes larger than the number of available degrees of freedom in the black hole, so that for times $t>t_{\rm Page}$   
the joint state of the black hole coupled to the radiation becomes mixed, thus contradicting the hypothesis B1 and giving rise to the paradox.

To avoid this problem and preserve unitarity, 
it was claimed that the entropy of the Hawking radiation should start decreasing at relatively early times when $t\sim t_{\rm Page}$ and vanish at the end of the evaporation~\cite{Page:1993wv,Page:1993up,Almheiri:2020cfm}. It is also sometimes claimed~\cite{Page:1993wv,Dvali:2015aja} that this might happen thanks to non-negligible non-thermal corrections arising  for times larger than $t_{\rm Page}.$  In this way the von Neumann entropy of the radiation would purify consistently with unitarity and follow the so-called Page curve; see Fig.~\ref{fig3}.
A price to pay in this scenario is the need for new physics far from the Planckian regime, i.e. abandoning of B2.

\subsection{Case C: No paradox}\label{case-C}

In the Case B we noticed that the assumption B3 plays a crucial role in giving rise to paradoxical conclusions even when quantum-gravity effects are still negligible, thus implying either the lack of a unitary evolution or the need for new physics already at early times when the effective-field-theory description should still be reliable. 

We now discuss a third scenario that is usually less appreciated, in which the assumption B3 is dropped and the only hypothesises are the following:

\begin{enumerate}
	
	\item[C1.] Quantum states evolve in a unitary way. In particular, pure states evolves into pure states;
	
	\item[C2.] Semiclassical general relativity is a valid low-energy effective field theory to describe black hole physics far from the Planckian regime.
	
\end{enumerate}	
These two assumptions do \textbf{not} necessarily lead to any information loss paradox! Since the thermodynamic entropy of the black hole is not constrained to be given by the horizon area, the outgoing Hawking quanta can always be entangled with the ingoing ones. In other words, the (negative energy) ingoing flux of the radiation contributes to the increase of the thermodynamic black hole entropy,  so that although the entropy of the radiation can become larger than the Bekenstein-Hawking entropy, i.e. $S_{\rm rad}>S_{\rm BH},$ it will never exceed the total black hole entropy $S_{\rm bh}$, and {\it no} entropy paradox arises.

It is worth mentioning that in analog gravity models for black holes~\cite{Barcelo2005} correlations between ingoing and outgoing quanta can be experimentally measured~\cite{Steinhauer:2015saa,Weinfurtner:2019zyc}.
Furthermore, results towards the resolution of the information problem in this context seem to suggest that the outgoing Hawking radiation is entangled even after the disappearance of the analog black hole~\cite{Liberati:2019}. The extrapolation of these partial results to  gravitational black holes seems to agree with this third scenario for which B3 does not hold. While it is true that the analogy is far from perfect, it is also true that the lessons drawn in this context should not be completely ignored as analog systems represent the closest we can get to the experimental detection of Hawking radiation.

Given the validity of the assumptions C1 and C2, we can surely state that the whole dynamics remains unitary as long as time scales shorter than the black hole life time are considered. Instead, when approaching the end point of the evaporation process quantum-gravity effects must be taken consistently into account, and to do so a full theory of quantum gravity is necessary.  

Therefore, it is important to stress that the validity of semiclassical general relativity before the black hole reaches Planckian size is \textbf{not} necessarily incompatible with the unitarity of quantum mechanics. It is important to stress that the information may or may not be lost in the final stages of the evaporation depending on the physics describing that regime, but this would depend on the full theory of quantum gravity and goes beyond the scope of semiclassical gravity. While this should definitely be a well-known fact, the information loss paradox is often stated as the incompatibility between  C1 and C2, whereas the hypothesis B3 is usually implicitly assumed and taken for granted. 

In fact, it should be emphasized that the information loss problem related to the growing of the von Neumann entropy is just a consequence of imposing B3 as a hypothesis. In other words, the imposition of the central dogma implies the emergence of a paradox which, otherwise, would not arise.
With this in mind, in the next section we carefully and critically inspect the motivations that are usually given in support of the central dogma.

\section{Standard arguments for the central dogma}
\label{Sec:arguments}

In the previous section we have shown that the requirement of the central dogma in the Case B, i.e. the imposition of the assumption B3, directly leads to a paradoxical conclusion. Therefore, we believe it is of crucial importance to better understand the motivations supporting this hypothesis.  

\begin{enumerate}

\item[(i)] To our understanding, the main motivation comes from the fact that black holes obey standard thermodynamic laws~\cite{Bardeen:1973gs} with an entropy given by the Bekenstein-Hawking formula~\cite{Bekenstein:1972tm,Bekenstein:1973ur}
\begin{equation}
S_{\rm bh}=S_{\rm BH}=\frac{A}{4G}.
\end{equation} 
From this observation, and from the fact that for a generic system the thermodynamic  entropy $S_{\rm th}$ imposes an upper bound on the von Neumann entropy $S_{\rm vN},$ i.e.
\begin{equation}\label{eq:S-comp}
    S_{\rm th}\geq S_{\rm vN},
\end{equation}

one would expect that the apparent-horizon area $A=16\pi G^2M^2$ has an intrinsic statistical nature and the total number of internal states might be bounded by $e^{S_{\rm BH}}.$

\item[(ii)] This picture is reinforced by the imposition of the Bekenstein bound~\cite{Bekenstein:1980jp}, which states that the entropy of any quantum system localized in a region of circumferential radius $R$ and of total energy $E$ is bounded as 
\begin{equation}
    S\leq 2\pi ER\,.\label{bek-bound}
\end{equation}
In the case of a Schwarzschild black hole we have $R=2GM$ and $E=M,$ so that the previous inequality reads
\begin{equation}
    S_{\rm bh}\leq \frac{A}{4G}=S_{\rm BH}\,.\label{bek-bound-BH}
\end{equation}
Therefore, if we assume the Bekenstein bound to be valid, then the Bekenstein-Hawking entropy is the maximal possible entropy for a black hole as seen from the outside. 

The Bekenstein bound seems to be not applicable when self-gravity effects are not negligible, e.g. for instance during a gravitational collapse and for sufficiently large cosmological regions. For this reason, Bousso proposed the so-called covariant entropy bound~\cite{Bousso:1999xy} whose range of applicability is wider than Bekenstein bound and reduces to the latter when self-gravity effects are negligible. For a Schwarzschild black hole the Bousso bound coincides with the one in Eq.~\eqref{bek-bound-BH}.

\item[(iii)] The previous bounds on the entropy seem to suggest an intrinsic connection between geometry and information. This has brought many people to believe in the existence of a holographic principle~\cite{tHooft:1993dmi,Susskind:1994vu} according to which the area of any hypersurface poses a limit on the information that can be stored within the adjacent spacetime regions. It is often claimed that holography should be a property of a consistent theory of quantum gravity~\cite{Bousso:2002ju}, and it can be considered the main modern motivation in support of the central dogma.

\item[(iv)] A more recent argument related to the previous ones comes from a computation of Hawking radiation that involves the use of the Ryu-Takayanagi entropy formula~\cite{Ryu:2006bv,Lewkowycz:2013nqa} in the context of holography and AdS/CFT correspondence~\cite{Maldacena:1997re}. Through such a computation one can reproduce a behavior for the von Neumann entropy of the radiation that is compatible with the Page curve. This result was interpreted as a strong indication towards the validity of the central dogma~\cite{Almheiri:2020cfm}. 
It must be also emphasized that this result has been rigorously obtained in simplified settings like 2D black hole spacetimes~\cite{Penington:2019npb,Penington:2019kki,Almheiri:2019psf,Almheiri:2019hni,Almheiri:2019qdq}.

\end{enumerate}

In the next section we will present a thought experiment illustrating a simple physical configuration in which the central dogma does \textit{not} hold; in particular, we will  comment on the limitations of the previous arguments.

\section{General relativity or central dogma?}
\label{Sec:counterex}

\subsection{Physical configuration}

Le us assume the hypothesis B2 to be valid and consider a physical setup in which a spherically symmetric solar mass black hole is radiating Hawking quanta and, at the same time, also accreting matter. The type of matter falling into the black hole is in principle completely arbitrary, but we can surely assume that its quantum state is entangled with a second component, so that the total state of such additional matter degrees of freedom is pure\footnote{For a different type of matter, e.g. in a mixed state, the following considerations will be more involved but they can be repeated and shown to be still valid.}. For instance, we can imagine a device that generates two fluxes of matter entangled with each other: one falls into the black hole, whereas the other reaches $\mathscr{I}^+.$   Therefore,  while the total state of matter is still pure,  an observer at infinity would only observe the outgoing flux of matter and measure a mixed state.

We work in the regime in which the outgoing flux of Hawking radiation is carrying an energy per unit of time much smaller than the mass of the black hole, so that the adiabatic approximation is valid and we can implement Hawking computation to obtain the temperature of the black hole:
\begin{equation}
T=\frac{1}{8\pi G M}\,.
\end{equation}
Then, the mass loss is given by the Stefan–Boltzmann law,
\begin{equation}
\frac{{\rm d}M}{{\rm d}v}=- \sigma_{\rm SB} T^4 A=-\sigma_{\rm SB}\frac{16\pi}{\left(8\pi \right)^4}\frac{1}{G^2M^2}\,,
\end{equation}
where $\sigma_{\rm SB}$ is the Stefan–Boltzmann constant and $v$ is the ingoing Eddington–Finkelstein coordinate which provides a notion of time for an observer on $\mathscr{I}^-$~\cite{Wald:1984rg}.

\begin{figure}[t]
	\includegraphics[scale=0.45]{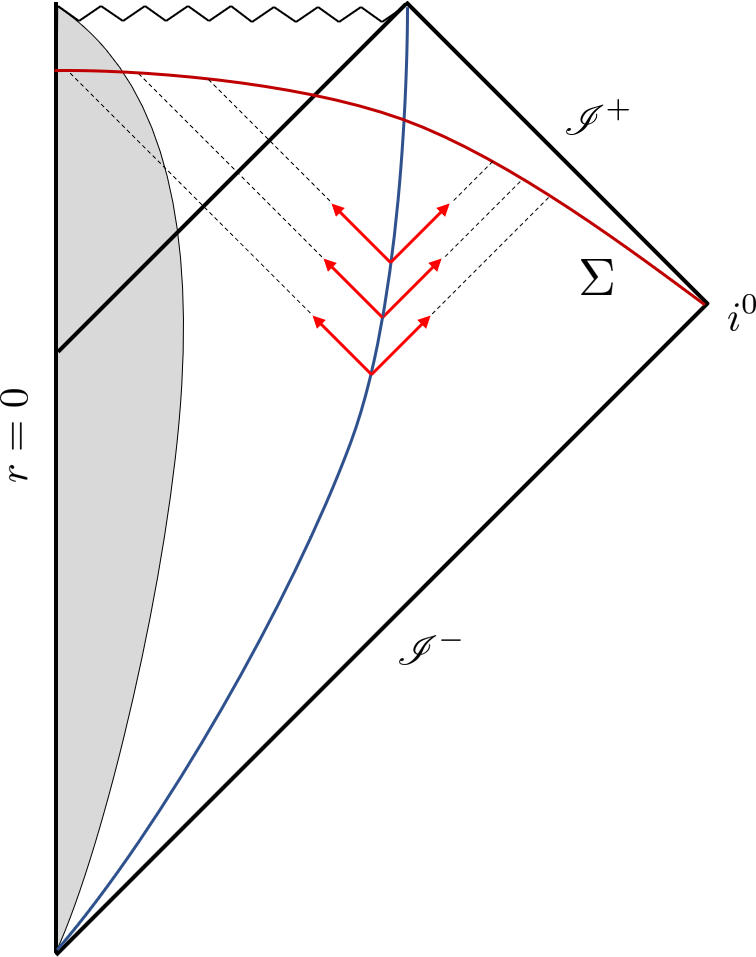}
	\centering
	\protect\caption{Penrose diagram of a black hole whose mass is much larger than Planck mass and that emits thermal radiation according to semiclassical general relativity. The blue solid line is a timelike hypersurface on which one places a device that emits two entangled fluxes of matter (red solid arrows): one falling into the black hole, while the other reaching $\mathscr{I}^+$. The black hole mass remains constant in time as it is assumed that the additional energy of the ingoing flux compensates the energy loss due to Hawking radiation.}
	\label{fig4}
\end{figure}

Furthermore, the ingoing flux of additional matter can be chosen arbitrarily but, for simplicity, we assume that it has the same energy as the outgoing flux of Hawking radiation. In this way, the mass of the black hole remains constant in time and, as a consequence, the horizon area (or, in other words, the Bekenstein-Hawking entropy $S_{\rm BH}=A/4G$) remains constant too.
The Penrose diagram for this physical setup is shown in Fig.~\ref{fig4}.  

We can now consider the entropy of the matter degrees of freedom for an asymptotic observer. If the semiclassical approach holds true in the regime under investigation (B2), then the horizon is smooth and we would not expect any kind of interaction between the outgoing Hawking radiation and the infalling flux of matter. Thus, the behavior of the von Neumann entropy $S_{\rm m}$ associated to the additional outgoing (or ingoing) matter is \textbf{independent} of whether $S_{\rm rad}$ follows Hawking computation or the Page curve. In fact, in both cases the entropy $S_{\rm m}$ is a monotonically increasing function of time; see Fig.~\ref{fig5} for a schematic illustration. This means that we do not need to require B1 to elaborate our argument.

Let us also point out that the states associated to the additional matter degrees of freedom are prepared outside the black hole, and therefore they can be engineered in such a way to be and stay orthogonal. This means that there is \textit{no} dependency relation among the matter states that could even in principle decrease the value of the entropy $S_{\rm m}.$

\subsection{Contradiction between B2 and B3 }
\label{subsec:contradiction-B2-B3}

The important point to notice in the above physical scenario is the following. If we assume the central dogma to be true, i.e. B3, then after some time  $S_{\rm m}$ would exceed the Bekenstein-Hawking entropy of the black hole and we would conclude that the total state of the additional matter is \textit{not} pure anymore. It is not surprising that we would obtain a similar conclusion as in the case of the entropy-problem version of the information loss paradox (i.e. Case B in Sec.~\ref{Sec:info paradox}), where in that case the role of $S_{\rm m}$ is played by $S_{\rm rad}.$


\begin{figure}[t!]
	\centering
	\subfloat[Subfigure 1 list of figures text][]{
		\includegraphics[scale=0.415]{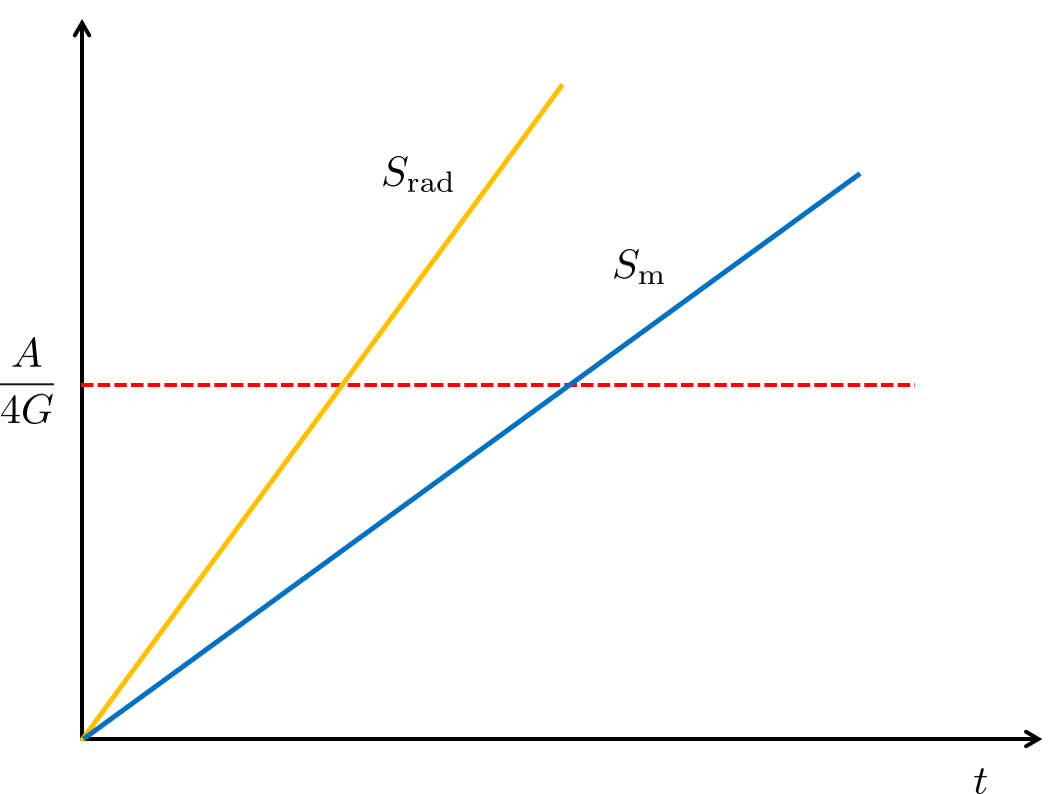}\label{fig5a}}\qquad
	\subfloat[Subfigure 2 list of figures text][]{
		\includegraphics[scale=0.415]{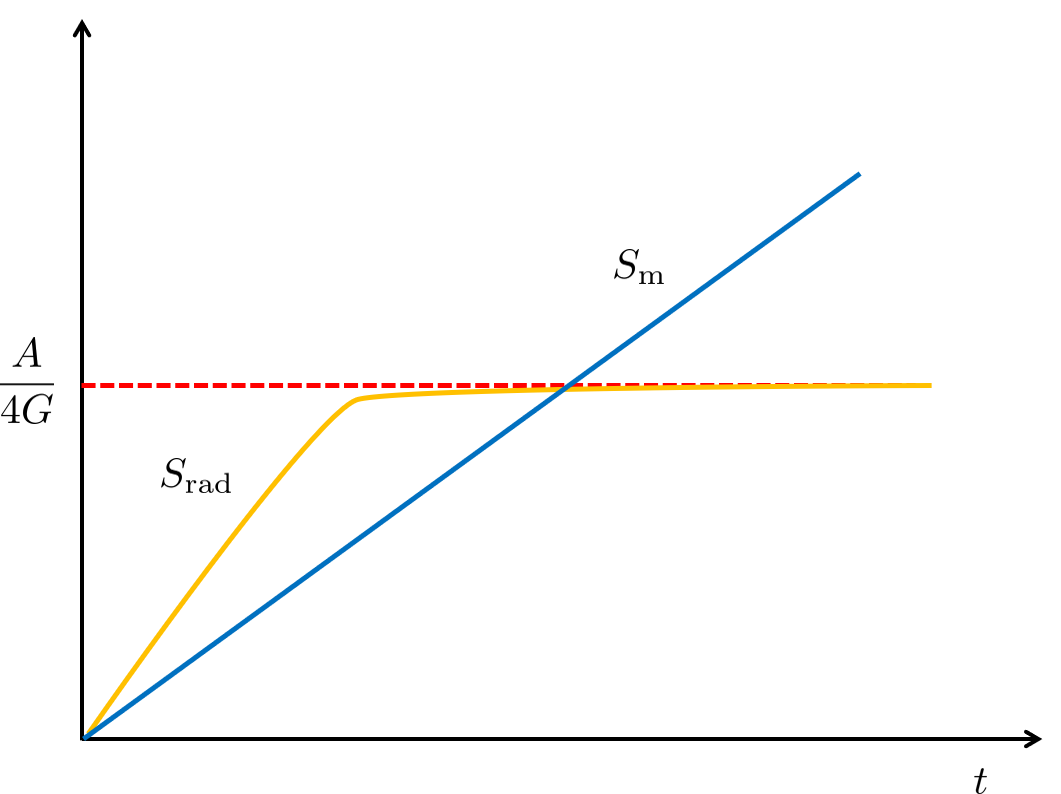}\label{fig5b}}
	\protect\caption{Schematic representations of the von Neumann entropies of the outgoing Hawking radiation $S_{\rm rad}$ (yellow solid line) and of the additional matter degrees of freedom $S_{\rm m}$ (blue solid line), in comparison with the gravitational Bekenstein-Hawking entropy $A/4G$ (red dashed line) which is assumed to be constant in our physical setup. In the plot (a) $S_{\rm rad}$ is determined by Hawking semiclassical computation; while in (b) $S_{\rm rad}$ follows the Page curve. In both cases, if we assume the central dogma to be true, then the number of degrees of freedom in the black hole is \textit{not} enough to maintain the entanglement between ingoing and outgoing matter fluxes. At the same time, the Hawking radiation and the additional matter cannot be entangled with each other if we assume general relativity to hold at the horizon scale.}
	\label{fig5}
\end{figure} 

However, in the setup just described everything is under control.  Indeed, we know for sure that the matter degrees of freedom inside the black hole are entangled with the degrees of freedom outside. In particular, since the pair of (ingoing and outgoing) fluxes is in an entangled pure state, at any instant of time (or, equivalently, on any Cauchy hypersurface) the degrees of freedom of matter (delivered by the ingoing flux) inside the black hole are the same number as the ones of matter (delivered by the outgoing flux) outside of it, and they will eventually exceed the maximum limit allowed by the central dogma, i.e. $S_{\rm m}>A/4G.$ This implies that the central dogma cannot be satisfied in our physical configuration. 

Remarkably, this conclusion does not depend on whether the evaporation process is unitary or not. Indeed our argument applies to both cases of Figs.~\ref{fig5a} and~\ref{fig5b}. Thus, without requiring B1, in our thought experiment we showed the existence of a contradiction between B2 and B3.

We strongly believe that the following fact should be highly emphasized.
The information loss problem is usually stated as an incompatibility between quantum mechanics and general relativity. However, the apparent issue only arises because of the imposition of the additional hypothesis B3 that is usually taken for granted; indeed, the usually claimed inconsistency is between B1 and $\text{B2}\cup\text{B3}$~\cite{Page:1993wv,Almheiri:2020cfm}. 

In fact, the information/entropy problem in Case B should be formulated as the incompatibility between the validity of semiclassical general relativity as a low-energy effective field theory far from the Planckian regime (implying a smooth horizon) and the central dogma. Therefore, we need to abandon either B2 or B3. In this respect, there exists \textit{no} information loss paradox since either $\text{B1}\cup\text{B2}$ (Case C) or $\text{B1}\cup\text{B3}$ does not necessarily lead to paradox.
In our opinion the assumption B3 should be critically revisited, and indeed the conclusion drawn from our thought experiment can be also phrased as a \textit{counterexample} to the central dogma if B2 is assumed to be the valid one.

Obviously, our argument would fail if some new physics at the horizon scale is invoked. However, since we expect quantum-gravity effects to emerge only when approaching the end of the evaporation, i.e. at Planck scales, then the effective-field-theory description is completely valid in the regime we considered. This is exactly the point that should be stressed. Contrary to what is usually stated, if \textit{no} new physics is assumed at the horizon scale then surely one needs to abandon B3 (i.e. to consider Case C) and then there is no information loss problem before reaching the latest stages of black hole evaporation where, instead, a consistent theory of quantum gravity is needed to make any prediction.

Finally, let us note that a physical setup with infalling additional matter similar to the one considered in our thought experiment was studied in \cite{Lowe1999}. The author argues that this configuration does not need to represent a counterexample to the central dogma if the degrees of freedom of the matter reach the singularity and are non-locally transferred to the radiation. This would lead to entanglement between the matter outside the black hole and the outgoing Hawking radiation. While this is certainly a possibility, it must be noted that the singularity is in the future of any observer. Therefore, the mechanism that would be necessary to transfer information from the singularity to the exterior would not only require interaction to propagate in a spacelike way (as pointed out in~\cite{Lowe1999}), but also backward in time. Furthermore, in our scenario this would require the outgoing matter to be maximally entangled  simultaneously with the ingoing matter and the outgoing Hawking radiation, thus violating the monogamy of entanglement theorem~\cite{Coffman:1999jd}.
Therefore, we believe that this is far from the conservative scenario as it would surely require new physics beyond the standard semiclassical description, thus violating B2. 

Let us also mention that in the context of quantum teleportation the information about a given quantum state can be sent combining a purely classical channel and a purely quantum one~\cite{Bennett:1992tv} and that the amount of violation of semiclassical causality might be significantly reduced~\cite{Mukohyama:1998xq}. Although the classical channel can in principle contain very little information, the quantum channel cannot be decoded by its own without the classical channel. Therefore, a mechanism that is able to teleport the quantum portion of the state would not give rise to causality violation. This partially alleviates the problem discussed in this section as all or a part of the information encoded in the quantum channel can leak out from the black hole without giving rise to any causality problem. However, unless at some point the classical channel is also transferred, the information in the quantum channel is useless. Invoking the transfer of the classical channel from the interior of a black hole to the exterior then requires violation of the semiclassical causality. For this reason, the full information can be recovered in accordance with the central dogma only if B2 is violated at least to the extent that allows for the transfer of the minimal amount of information in the classical channel.

\subsection{Stronger contradiction}\label{Sec:stronger_contr}

Through the thought experiment described in the previous subsection we argued that the entropy problem in Case B should be reformulated as a contradiction between B2 and B3 independently of B1, instead of an incompatibility between B1 and $\text{B2}\cup\text{B3},$ contrarily to what is often claimed. 

Before continuing our analysis, let us note that the assumption B2 can be split in two parts B1a and B2b such that $\text{B2}=\text{B2a}\,\cup\,\text{B2b}:$\footnote{It is worth mentioning that our assumptions B2a, B2b and B3 coincide with the postulate 2, 4, 3 in~\cite{Almheiri:2012rt}, respectively; whereas our assumption B1 differs substantially from postulate 1 of~\cite{Almheiri:2012rt} as we do not assume that the Hawking radiation must purify. Moreover, in~\cite{Almheiri:2012rt} no contradiction between B2b and B3 was noticed; in fact, B3 was implicitly assumed to be true regardless of the validity of B2b.} 
\begin{enumerate}
		
	    \item[B2a.] Black holes whose mass is larger than Planck mass emit thermal radiation according to semiclassical general relativity;
		
		\item[B2b.] Infalling matter far from the Planckian regime obeys the laws of general relativity.

\end{enumerate}
The hypothesis B2a means that the semiclassical approach can be trusted outside the horizon for time scales shorter than black hole life time.  Whereas, B2b means that in the presence of a black hole emitting radiation, an infalling matter far from the Planckian regime would experience nothing special, e.g. when crossing and leaving the horizon in the past, consistently with the equivalence principle, so that the horizon is assumed to be a smooth surface.

The contradiction we found in our thought experiment is between B2b and B3, that is, between the laws of general relativity governing the physics of infalling matter and the central dogma; while B2a and B3 can in principle be perfectly compatible. This means that the contradiction is even stronger. 

On the other hand, one could criticize our conclusion by saying that to formulate the paradox in Case~B only B2a is really needed, and the incompatibility is between B1 and $\text{B2a}\cup\text{B3},$ independently of B2b. Although assuming B2a without B2b might require some discontinuous jump in the physics governing the proximity of the horizon, we agree that such a logic can be correct and can lead to an information loss paradox. However, we should immediately notice that this does \textbf{not} imply an incompatibility between general relativity and quantum mechanics since abandoning B2b already implies something beyond general relativity.

Indeed, in our thought experiment we clearly showed the existence of a contradiction between the assumptions B2b and B3, from which it follows that if B3 is valid, then the infalling matter should experience some new physics whose description goes beyond the laws of general relativity. Remarkably, this implies that the information loss problem in the Case~B (with B1, B2a, B3) is a statement of an inconsistency between quantum mechanics and some new unknown gravitational physics incompatible with general relativity. 
In other words, this version of Case B does not show any contradiction between general relativity and quantum mechanics. 

Furthermore, if  we assume B2b,
then we must drop B3 and we fall into Case C from which it is clear that \textit{no} information paradox arises before reaching the Planck scale where, instead, a consistent theory of quantum gravity is needed to make any prediction about the black hole final state. Therefore, we should say that, in the case B, \textit{the real incompatibility is between general relativity and  the information loss paradox itself.}

\subsection{Limitations of the standard arguments}

One way to interpret the conclusion reached in our thought experiment is that we provided a simple working example in which the central dogma does \textit{not} hold (if B2b is true). Let us now analyze where the usual arguments in favor of the central dogma reviewed in Sec.~\ref{Sec:arguments} fail.
\begin{enumerate}

    \item[(i)] The standard thermodynamic argument based on the area-law for the entropy of a black hole fails because it does not take into account the contribution due to the matter fields inside the black hole, e.g. $S_{\rm m}$ in our thought experiment. Indeed, it should be clarified that the black hole thermodynamic laws~\cite{Bardeen:1973gs} are intrinsically gravitational in nature. In other words, the Bekenstein-Hawking entropy $S_{\rm BH}$ takes into account only gravitational degrees of freedom, and in general the total black hole entropy is given by $S_{\rm bh}=S_{\rm BH}+S_{\rm m}.$
    
    To understand why additional matter degrees of freedom must be taken into account for the correct evaluation of the von Neumann entropy, despite they do not appear in the gravitational thermodynamic laws (thus apparently violating the inequality \eqref{eq:S-comp}), it is useful to consider a system composed by two subsystems ``$A$'' and ``$B$''. If these two subsystems are very far apart, it is possible to study the thermodynamic laws of subsystem $A$ without taking into consideration subsystem $B$. However, this is no longer true if we are interested in the evaluation of the von Neumann entropy, indeed in this case we need to consider both subsystems even if they are very far apart because of quantum correlations due to entanglement. Eq.~\eqref{eq:S-comp} is only valid if we consider the thermodynamic entropy of the whole system $A\cup B,$ i.e.
    \begin{equation}\label{eq:S-comp2}
        S_{\rm th}(A\cup B)\geq S_{\rm vN}(A\cup B)\,,
    \end{equation}
    but obviously it does not hold if we consider only the thermodynamic entropy of the subsystem $A$, i.e.
    \begin{equation}
        S_{\rm th}(A)\ngeq S_{\rm vN}(A\cup B)\,.
    \end{equation}
    While this consideration may seem obvious, it is at the root of the contradiction reached in our thought experiment.
    The role of subsystem $A$ is played by the gravitational degrees of freedom, whereas subsystem $B$ is given by the matter degrees of freedom that fall into the black hole. An outside observer cannot be influenced by such matter degrees of freedom, thus they do not need to be included in the description of black hole thermodynamics. However, if these matter degrees of freedom are entangled with degrees of freedom outside, as it happens in our setup and with Hawking radiation, then we cannot discard them when computing the von Neumann entropy. In particular, the fact that at the Page time the entropy of the matter outside (or inside) the black hole exceeds the area is \textbf{not} in contradiction with Eq.~\eqref{eq:S-comp2}, but it simply confirms  that the apparent-horizon area only accounts for the entropy associated to the gravitational degrees of freedom. Thus, the inequality
    \begin{equation}
        S_{\rm vN}\left( \text{gravity}\cup \text{matter} \right)>\frac{A}{4G}
    \end{equation}
for some states does \textbf{not} lead to any conceptual problem, rather it implies
    \begin{equation}
      \frac{A}{4G}=S_{\rm th}\left(\text{gravity}  \right)\neq S_{\rm th}\left(\text{gravity}\cup\text{matter}  \right).
    \end{equation}
Therefore, a priori there is no reason why the addition of matter degrees of freedom should not increase the entropy of the black hole, especially if B2 is valid. The same conclusion would also apply to the ingoing Hawking flux which should contribute to an increase of the thermodynamic entropy of the entire black hole system made up of gravitational plus matter (Hawking quanta) degrees of freedom.

The possibility that the Bekenstein-Hawking entropy only accounts for the degrees of freedom that are accessible (i.e. causally connected) to an outside observer was already suggested in \cite{Rovelli:2019tbl}. Our thought experiment allow us to conclude that this must be the case if the assumption B2b is satisfied.
    
    This analysis can also easily explain the apparent contradiction between the Bekenstein-Hawking formula and the fact that ``bag of gold'' spacetimes~\cite{Wheeler:1964qna} can have a huge volume and a very small area. The area law simply does not describe all the bulk degrees of freedom inside a black hole. See also Refs.~\cite{Christodoulou:2014yia,Rovelli:2017mzl} for a complementary argument against the central dogma.

    \item[(ii)] It is often stated that the ingoing flux of Hawking radiation  cannot increase the entropy of the black hole otherwise the Bekenstein bound in Eqs.~\eqref{bek-bound} and~\eqref{bek-bound-BH} would be violated. However, it must be emphasized that the standard Bekenstein bound cannot be naively applied when both negative and positive energies are involved, and it is not clear whether gravitational corrections can appear when the curved nature of spacetime is taken into account. In fact, a version of such a bound has been rigorously proven only in the context of quantum field theory in flat spacetime~\cite{Casini:2008cr}. To be more precise, although the proved inequality is applicable to any quantum systems and to any quantum states, it can be interpreted as Bekenstein bound only in special cases.

    From these last observations, we also see no reason why the infalling matter in our thought experiment should not contribute to the increase of the black hole entropy and not violate the Bekenstein bound. 
    Moreover, the Bousso entropy bound, that for a black hole implies the one in Eq.~\eqref{bek-bound-BH}, would also be violated. Indeed, both the statement of the Bousso bound (the covariant entropy conjecture) and its spacelike projection assume the dominant energy condition~\cite{Bousso:1999xy} that does not hold in quantum field theory in curved spacetime. 

    \item[(iii)] As we mentioned in Sec.~\ref{Sec:arguments}, holography is one of the main motivation that is usually advocated in support of the central dogma. However, our thought experiment simply reveals an incompatibility between B2b and B3, where the latter is normally thought to be a necessity in presence of holography. Therefore, from our argument it follows that holography, if it implies B3, and the validity of semiclassical gravity for matter infalling to a black hole would be incompatible.
    
We are not claiming that holography is wrong but argued that it is incompatible with the low-energy effective-field-theory approach according to which the horizon is smooth as long as the black hole mass is larger than Planck mass. For instance, quantum-gravity theories that predict new physics at the horizon scale and that invalidate B2b might be consistent with a holographic principle~\cite{Bousso:2002ju}. 
    However,  we do not see any strong reason why quantum gravity should unconditionally incorporate or predict holography. There may be a condition under which holography holds and such a condition may be violated in some cases. Indeed, it is worth mentioning that several recent approaches~\cite{Tomboulis:2015esa,Anselmi:2017ygm,Anselmi:2018ibi,Donoghue:2018izj,Salvio:2018crh,Holdom:2021hlo,Percacci:2017fkn,Reuter:2019byg,Platania:2018eka,Bonanno:2020bil} aimed at formulating a renonormalizable and unitary quantum field theory of the gravitational interaction seem to be valid proposals for a consistent quantum-gravity theory that does not rely on or have any relation with holography.
    
    \item[(iv)] Very recently, many interesting works have been done towards the resolution of the entropy problem outlined in Case B~\cite{Penington:2019npb,Penington:2019kki,Almheiri:2019psf,Almheiri:2019hni,Almheiri:2019qdq}. Although we appreciate that rigorous computations of the Page curve have been made in some specific simplified settings, we have reservations about the claimed interpretation of the result. Let us briefly review the logic that was followed.

    The computation of the von Neumann entropy of the radiation was obtained by using the Ryu-Takayanagi formula~\cite{Ryu:2006bv} that can be derived from the gravitational path integral under the assumption of holography~\cite{Lewkowycz:2013nqa,Harlow:2020bee}. As a result the Page curve was reproduced, and several authors~\cite{Penington:2019npb,Penington:2019kki,Almheiri:2019psf,Almheiri:2019hni,Almheiri:2019qdq} claimed that the entropy problem (paradox) was solved.  Let us also emphasize that the reason why the final state of the radiation purifies in their calculation is that the Ryu-Takayanagi formula for the entropy of the radiation  also includes  the contribution from the black hole interior.  This approach to compute entropy of the radiation is also known as ``island program''~\cite{Almheiri:2020cfm}.

     One may now wonder what would be the outcome of the island computation when applied to the von Neumann entropies of radiation and additional matter in our thought experiment. In fact, if one assumes that the Hilbert space of the interior of the black hole (including the additional matter) is no longer independent of the Hilbert space of outgoing radiation, then by using the island formula one could show that eventually the quantity $S_{\rm rad}+S_{\rm m}$ will tend to $A/4G$ and never exceed it. If this was the case, then one would conclude that the central dogma is still respected.

     We agree with this possibility; however, we believe that such a scenario is completely consistent with the conclusion drawn from our thought experiment. The assumption B3 of central dogma can be satisfied only if B2b is violated. Indeed, in our physical setup the island would correspond to the entire black hole interior, and it so happens that the von Neumann entropy $S_{\rm m}$ of the infalling matter does not contribute to the von Neumann entropy of the black hole because both ingoing and outgoing fluxes of matter are taken to be part of the same $\text{island}\cup\text{radiation}$ region. This fact implicitly assumes that the additional matter and the radiation get somehow entangled, but this can happen \textit{if and only if} the matter in free fall interacts with the Hawking radiation, which would be impossible if the laws of general relativity are valid, i.e. if B2b holds.
    
    Therefore, when applied to our physical setup the island program implies that the laws of semiclassical general relativity to describe infalling matter are violated. This might also suggest that new physics beyond general relativity is needed to completely trust and physically interpret the island computation\footnote{Also, in our opinion it is not totally clear which Euclidean
solutions should contribute to the island computation based on the
replica method as saddle points, without additional information or
assumptions.}. 
Yet another possible interpretation is that the island formula somehow takes into account the transfer of information through the quantum channel in the language of quantum teleportation. However, as explained in the last paragraph of subsection \ref{subsec:contradiction-B2-B3}, one still needs to transfer the classical channel from the black hole interior to the exterior and thus the semiclassical general relativity and the central dogma are incompatible with each other. 

See also Ref.~\cite{Geng:2021hlu} where it was recently argued that the island program/holography might not be consistent with massless gravitational theories, and Ref.~\cite{Omiya:2021olc} where it was claimed that the transfer of information based on the island computation requires a nonlocal interaction term in the Hamiltonian. On the other hand, it is also worth mentioning that other authors~\cite{Harlow:2020bee} claim that holography must be imposed even when performing the standard Euclidean computation~\cite{Gibbons:1976ue} of the Gibbons-Hawking entropy, or in other words that the Euclidean gravitational path integral is only valid for holographic theories.

\end{enumerate}

\section{On the microscopic interpretation of $A/4G$}\label{Sec:micr}

Let us now assume that general relativity is a valid theory far from the Planckian regime, so that the horizon is a smooth surface and a free falling observer does not experience anything out of the ordinary before approaching the singularity. Consequently, the central dogma (B3) is violated, and we would fall into Case C discussed in Sec.~\ref{case-C} according to which no information paradox exists. In such a case, a very natural question to ask is -- what is the microscopic interpretation of the Bekenstein-Hawking entropy $A/4G?$

Macroscopically, i.e. in a coarse-grained sense, $A/4G$ can be interpreted as the contribution of a black hole to the total thermodynamic entropy that follows the generalized second law in semiclassical regimes. However, since the central dogma does not hold, then one should give an alternative interpretation of the Bekenstein-Hawking formula at microscopic level, i.e. one should explain which part of the black hole degrees of freedom are described by $A/4G.$ or instance, in Ref.~\cite{Frolov:1993ym} it was assumed that the dynamical degrees of freedom of the black hole correspond to thermally excited modes behind the horizon that are invisible to a distant observer, and it was shown that their contribution is indeed proportional to the horizon area.

 One expects the microscopic description to be highly model-dependent, and that the correct origin and counting of both gravitational and matter degrees of freedom can be given by a consistent theory of quantum gravity. In what follows we address the above question in three different well-known theories/scenarios.
\paragraph{String theory:}In Ref.~\cite{Strominger:1996sh} it was shown that the counting of D-brane states correctly reproduces the formula $A/4G$ for a class of five-dimensional extremal black holes. In particular, no more entropy than $A/4G$ can be added to the D-brane, which is supposed to correspond to a black hole. Furthermore, such correspondence may be more general and applicable to realistic non-extremal black holes: the gravitational radius of a D-brane increases as the string coupling is raised; and the D-brane becomes a black hole when the gravitational radius becomes as large as the size of a string. If the discontinuity in the mass at the transition is not too large and if the typical D-brane states become the typical black hole states then the entropy of the D-brane correctly accounts for the entropy of the black hole up to a numerical factor of order unity~\cite{Horowitz:1996nw}. Then, apparently, the black hole entropy cannot significantly exceed the Bekenstein-Hawking value. However, even if the computation can be extended to realistic non-extremal black holes, we should notice that it is not known which degrees of freedom in the black hole spacetime correspond to the D-brane; they may be degrees of freedom localized at the horizon, those localized slightly inside or outside the horizon, those localized near the singularity of the classical black hole solution, or something else. If the D-brane corresponds to degrees of freedom localized or extended outside the horizon in the black hole spacetime then B2b may be violated. Whereas, if the D-brane corresponds to some degrees of freedom localized somewhere strictly inside the horizon (e.g. near the classical singularity), then additional entropy may be stored in a region inside the horizon but away from the place where those degrees of freedom corresponding to the D-brane are localized. In this case, one might argue that the D-brane states would correspond to the gravitational degrees of freedom, i.e. $A/4G$, and that the additional states would correspond to the matter degrees of freedom, i.e. $S_{\rm m}.$ Thus in this case, consistently with our thought experiment, the total entropy of the black hole would be given by $S_{\rm bh}=A/4G+S_{\rm m}>A/4G$. In summary the D-brane picture of black holes in the context of string theory has not yet been fully understood and therefore it is at this stage compatible with various possibilities, such as the one suggested by our thought experiment.

\paragraph{Loop quantum gravity:} In Ref.~\cite{Ashtekar:1997yu} the Bekenstein-Hawking formula for the entropy was computed in the non-perturbative framework of loop quantum gravity. In particular, by making a quantized phase-space description for the black hole, it was shown that the resulting statistical mechanical entropy is proportional to the horizon area; see also Refs.~\cite{Rovelli:1996dv,Meissner:2004ju} for related works. The important point to notice in the derivation is that the computed entropy only accounts for the quantum states that describe the horizon geometry, as also pointed out in Ref.~\cite{Ashtekar:1997yu}. Therefore, also in this approach the horizon area counts purely gravitational degrees of freedom; whereas, matter degrees of freedom are expected to give an additional contribution to the total entropy of the black hole (as seen from outside).

\paragraph{Corpuscular gravity:} According to the corpuscular picture~\cite{Dvali:2011aa,Dvali:2012en,Dvali:2013eja,Dvali:2014ila}, a black hole can be described as a self-sustained system of weakly interacting gravitons. In this case, the number of gravitons $N$ coincides with the Bekenstein-Hawking formula, i.e. $N\simeq A/4G.$ Therefore, the gravitons microscopically count the gravitational degrees of freedom of the black hole. Whereas, if we also include matter in the system, then the corresponding degrees of freedom are associated with an additional (non-gravitational) entropy contribution, i.e. $S_{\rm m},$ so that the total black hole entropy is given by $S_{\rm bh}\simeq N+S_{\rm m}>A/4G.$ It is worth mentioning that, strictly speaking, for a corpuscular black hole no geometric notion of horizon exists so that, in general, also B2b is violated. However, one can show that in the $N\gg 1$ limit the corpuscular corrections are negligible, and the values of all relevant physical quantities are compatible with the presence of a smooth horizon~\cite{Dvali:2011aa}.

\section{Comments on black hole complementarity and firewall}\label{Sec:Comp-fir}

Let us now make some comments on black hole complementarity~\cite{Susskind:1993if} and firewall paradigm~\cite{Almheiri:2012rt}. In both cases, the central dogma is assumed as a postulate and never questioned.

\paragraph{Black hole complementarity:} According to black hole complementarity~\cite{Susskind:1993if,Susskind:1993mu}, an outside observer can replace the black hole in terms of a hot membrane whose surface lays one Planck length above the horizon at the so-called stretched horizon, and whose entropy coincides with the Bekenstein-Hawking  entropy $A/4G$ in agreement with the central dogma. Whereas, an infalling observer would not see any membrane in the proximity of the horizon, and would not experience any deviation from general relativity. 

However, from the conclusion of our thought experiment, it follows that the entropy of the stretched horizon (i.e. of the black hole) can be bounded by the area only if some new physics beyond general relativity is invoked at the horizon scale; otherwise an outside observer can prepare an experiment and confirm that the actual entropy is larger than $S_{\rm BH}.$ Therefore, the postulates of black hole complementarity are \textit{not} compatible with a smooth horizon.

\paragraph{Firewall:} The firewall paradigm~\cite{Almheiri:2012rt} also states that black hole complementarity is not compatible with a smooth horizon. However, we should stress that our logic is completely different from the firewall one. Indeed, in the firewall proposal the central dogma was imposed as a postulate and, moreover, the state of Hawking radiation was assumed to purify. In our thought experiment we have not made any assumption on the final state of the Hawking radiation; in fact, our conclusions are more general. We showed the existence of an incompatibility between the validity of general relativity for infalling matter and the central dogma, independently of whether the von Neumann entropy of the Hawking radiation follows the Page curve or not.

Furthermore, regarding the question of whether a firewall exists or not, one cannot really answer without having a consistent theory of quantum gravity which would allow us to predict the final quantum state of black hole evaporation. However, we do not see a strong reason why the state of the Hawking radiation should purify. For instance, if we abandon the central dogma, then it is also very plausible that the final black hole state is a remnant, so that the Hawking radiation remains in a mixed state but the total state of the joint system is pure; see also Ref.~\cite{Rovelli:2019tbl} for a similar argument.

\section{Discussion}\label{Sec:discussions}

In this paper, we critically inspected the assumptions behind the formulation(s) of the information loss problem.
Although it is often stated as an incompatibility between general relativity and the unitarity of quantum mechanics, we have argued that this is true only if the semiclassical gravity description is naively assumed to be valid all the way up to the Planck scale, and if the spacetime is extended beyond the Cauchy horizon in a specific way (Case A). Therefore, this version of the paradox is not particularly worrisome as quantum-gravity effects are expected to become relevant at the final stages of the black hole evaporation.

However, an ``apparent'' paradoxical conclusion involving general relativity and quantum mechanics is claimed to arise in regimes where the semiclassical approach is still valid, once the so-called central dogma is taken as an additional hypothesis (Case B). 

We questioned the well-posedness of this second formulation of the paradox. Remarkably,  by working in a well-defined physical setup consisting of a black hole whose energy loss due to radiation emission is compensated by an additional infalling flux of matter, 
we provided a clear working example to show the existence of an incompatibility between the following two assumptions:
\begin{itemize}
		
		\item[B2b.] Infalling matter far from the Planckian regime obeys the laws of general relativity;
		
		\item[B3.] As seen from the outside, a black hole behaves like a quantum system whose number of degrees of freedom is given by $A/4G,$ with $A$ being the apparent-horizon area.

\end{itemize}

As a consequence, we argued that Case B does not imply any contradiction between general relativity and quantum mechanics. Indeed, if the entropy problem is formulated as in Sec.~\ref{CaseB}, then the incompatibility is between the validity of semiclassical gravity far from the Planckian regime and the central dogma, independently of the assumption of unitary evolution, rather than between the assumption of unitary evolution and semiclassical gravity, contrarily to what it is often claimed~\cite{Almheiri:2020cfm}. 

On the other hand, if the semiclassical gravity assumption B2 is split into B2a and B2b as described in Sec.~\ref{Sec:stronger_contr}, and if we abandon B2b, then the paradox can be reformulated as an incompatibility among B1, B2a and B3. However, this reformulation must require the violation of B2b, which means that it is
automatically inconsistent with the laws of general relativity to describe infalling matter. Therefore, the paradox would consist in an incompatibility between quantum mechanics and some new unknown gravitational physics at the horizon scale. 
Furthermore, it seems at the very least unnatural to assume the validity of semiclassical gravity outside the horizon while simultaneously violating general relativity at the horizon scale.

We are now in a position to address the two questions raised in the titles of Sec.~\ref{Sec:info paradox} and Sec.~\ref{Sec:counterex}.
\begin{enumerate}

\item \textbf{Is there an information loss paradox?}\\
Very interestingly, our conclusions imply that \textit{the usually stated information loss paradox has nothing to do with the loss of information}, as quantum mechanics and semiclassical general relativity can be
perfectly compatible before reaching Planck scales where, instead, a consistent theory of quantum gravity is needed to make any prediction about the final state of black hole evaporation.
A contradiction only arises once the central dogma (B3) is added to the picture. The main message of this paper is that the problem is \textit{not} due to an incompatibility between general relativity and quantum mechanics but, instead,  it is due to a contradiction between B2b and B3 (and so between B2 and B3) independently of B1.
To be fair, we should say that,  in the case B, \textit{the real incompatibility is between general relativity and  the information loss paradox itself.}

Information may or may not be lost due to the formation of an event horizon~\cite{Visser:2014ypa}. However, here we have shown that there is no reason to argue in any direction within the regime of validity of semiclassical gravity.

\item \textbf{General relativity or central dogma?}\\
While it is crucial to properly formulate the problem, the question of which of the two assumptions (B2b or B3) should be dropped cannot be answered. In fact, there are a number of approaches that aim to address the process of black hole evaporation; 
for an incomplete list see~\cite{Almheiri:2020cfm,Penington:2019npb,Penington:2019kki,Almheiri:2019psf,Almheiri:2019hni,Almheiri:2019qdq,Marolf:2020rpm,Dvali:2011aa,Dvali:2012en,Susskind:1993if,Visser:2014ypa,tHooft:1996rdg,Betzios:2016yaq,Gaddam:2020mwe,Mathur:2005zp,Hayward:2005oet,Giddings:2012gc,Hawking:2014tga,Hawking:2016msc,Frolov:2014jva,Frolov:2017rjz,Bardeen:2014uaa,DAmbrosio:2020mut} and references therein. 
However, we should emphasize that it is impossible to understand whether one of these possibilities is indeed physically correct as it is currently impossible to perform an experiment to observe Hawking evaporation.
Therefore, the reader is free to choose which of the two assumptions to save, but at most one.

\end{enumerate}
%


\subsection*{Acknowledgements}
The authors would like to thank Gia Dvali, Valeri Frolov, Stefano Liberati, and Tadashi Takayanagi for critical comments and useful suggestions on an initial draft of this manuscript. In particular, they are grateful to Tadashi Takayanagi for enlightening discussions. F.~D.~F.~would like to thank Ibrahim Akal,  Ra\'ul Carballo-Rubio, and Mohammad Ali Gorji  for several stimulating discussions. L.~B.~is grateful to YITP for the warm hospitality, where this work was started.
L.~B.~acknowledges financial support from JSPS and KAKENHI Grant-in-Aid for Scientific Research No.~JP19F19324.
The work of F.~D.~F.~was supported by Japan Society for the Promotion of Science Grant-in-Aid for Scientific Research No.~17H06359.
The work of S.~M.~was supported in part by Japan Society for the Promotion of Science Grants-in-Aid for Scientific Research No.~17H02890, No.~17H06359, and by World Premier International Research Center Initiative, MEXT, Japan.

\bibliographystyle{utphys}
\bibliography{References}

\end{document}